\documentclass[aps,
  prl,
  nofootinbib,
  unsortedaddress,
  superscriptaddress,
  reprint,
  twocolumn]{revtex4-1}

\usepackage{amsmath,amssymb}
\usepackage{graphicx}
\graphicspath{{figures/}}

\newcommand{\figref}[1]{Fig.~\ref{#1}}
\newcommand{\Figref}[1]{Figure~\ref{#1}}
\newcommand{\ie}{i.e.,\,}
\newcommand{\mySec}[1]{{\bf #1}\,---}

\begin{document}
\title{A New Semiclassical Picture of Vacuum Decay}

\author{Jonathan Braden}
\affiliation{Department of Physics and Astronomy, University College London, Gower Street, London, WC1E 6BT, UK}
\author{Matthew C.\ Johnson}
\affiliation{Department of Physics and Astronomy, York University, Toronto, ON, M3J 1P3, Canada}
\affiliation{Perimeter Institute for Theoretical Physics, 31 Caroline St.\ N, Waterloo, ON, N2L 2Y5, Canada}
\author{Hiranya V.\ Peiris}
\affiliation{Department of Physics and Astronomy, University College London, Gower Street, London, WC1E 6BT, UK}
\affiliation{The Oskar Klein Centre for Cosmoparticle Physics, Stockholm University, AlbaNova, Stockholm, SE-106 91, Sweden}
\author{Andrew Pontzen}
\affiliation{Department of Physics and Astronomy, University College London, Gower Street, London, WC1E 6BT, UK}
\author{Silke Weinfurtner}
\affiliation{School of Mathematical Sciences, University of Nottingham, University Park, Nottingham, NG7 2RD, UK}
\affiliation{Centre for the Mathematics and Theoretical Physics of Quantum Non-Equilibrium Systems, University of Nottingham, Nottingham, NG7 2RD, UK}

\begin{abstract}
We introduce a new picture of vacuum decay which, in contrast to existing semiclassical techniques, provides a real-time description and does not rely on classically-forbidden tunneling paths. Using lattice simulations, we observe vacuum decay via bubble formation by generating realizations of vacuum fluctuations and evolving with the classical equations of motion.  The decay rate obtained from an ensemble of simulations is in excellent agreement with existing techniques. Future applications include bubble correlation functions, fast decay rates, and decay of non-vacuum states.
\end{abstract}
\maketitle

\mySec{Introduction}
Field theories with complex interaction potentials arise in a diverse range of physical applications: from cosmological theories of the multiverse inspired by string theory~\cite{Bousso:2000xa,Susskind:2003kw}, to cold atom Bose-Einstein condensates~\cite{Fialko:2014xba,Fialko:2016ggg,Braden:2017add}, to the dynamics of protein folding~\cite{doi:10.1146/annurev.physchem.48.1.545}.
Such theories generically possess many local potential minima.
At the homogeneous classical level such local potential minima are stable.
However, the uncertainty principle ensures that spatially-inhomogeneous quantum fluctuations are present, resulting in the metastablility of false vacuum field configurations.
The decay of false vacuum states is expected to proceed analogously to first-order phase transitions driven by statistical fluctuations in condensed matter systems~\cite{Langer:1967ax,Langer:1969bc}, with bubbles of a new phase nucleating then expanding into the ambient false vacuum.
The prevailing, if implicit, wisdom is that bubbles nucleate via non-classical paths in field space, analogous to tunneling through a barrier in quantum mechanics~\cite{Coleman:1977py,Callan:1977pt,Coleman:1977th,Lee:1985uv,Wainwright:2011kj,Weinberg:2012pjx,Greene:2013ida,Turok:2013dfa}. We propose an alternative picture of vacuum decay, in which the classical evolution of the field from some initial realization of the false vacuum fluctuations leads to the emergence of bubbles. As convenient nomenclature, we will refer to these as classically-forbidden and classically-allowed nucleations, respectively. 

We re-examine the nature of bubble nucleations during false vacuum decay in relativistic scalar field theory using semiclassical scalar field lattice simulations. 
In our framework, quantum effects are included by sampling realizations of the initial quantum state of the field, then time-evolving using the classical equations of motion.
As we demonstrate in the supplementary material, this captures the time-evolution of the wave functional to $\mathcal{O}(\hbar)$ (see also Ref.~\cite{Mrowczynski:1994nf} for a similar derivation). This treatment is analogous to the truncated Wigner approximation~\cite{PhysRev.40.749,QuantumNoise} often employed in atomic physics. 
We explicitly demonstrate the existence of classically-allowed paths connecting these initial realizations of the quantum false vacuum, to a subsequent state with nucleated bubbles of ``true vacuum''.
This demonstrates that false vacuum decay can proceed via classically-allowed dynamical evolution.
Furthermore, we compare the nucleation rates inferred using classical statistics for the initial fluctuations (\ie ignoring quantum interference), and find good agreement with instanton-based predictions which cannot tackle the time-dependent nature of the problem.
This suggests that the decay of the false vacuum proceeds via classically-allowed bubble nucleation events.
The supplementary material shows how this approach can be interpreted as an expansion of the full quantum dynamics in $\hbar$.

In this work, we focus on the particular field theory emerging from the cold atom analog false vacuum proposal of Fialko et al~\cite{Fialko:2014xba,Fialko:2016ggg,Braden:2017add}.
However, we do not expect the presence of classically-allowed decay channels to depend strongly on the specific choice of field theory.
The possible connection to experiment is particularly exciting in this context, as it provides a window to experimentally confirm the dynamics of vacuum decay, and possibly uncover additional effects arising from quantum interference such as Anderson localization~\cite{Anderson:1958vr,Podolsky:2007vg}.

\mySec{Semi-Classical Statistical Formalism}
We study relativistic false vacuum decay using semi-classical statistical lattice simulations.
Throughout we work in units with $\hbar = c = 1$.
In order to generate sufficient statistics to compute decay rates, we will work with fields in {$1+1$-dimensions}, although we do not expect consideration of higher spatial dimensions to qualitatively change our conclusions.

The fields are initialized as
\begin{equation}
  \phi({\bf x},t=0)       = \phi_{\rm fv} + \delta\hat{\phi}({\bf x}) \qquad 
  \dot{\phi}({\bf x},t=0) = \delta\hat{\dot{\phi}}({\bf x}) \, ,
\end{equation}
where the fluctuations $\delta\hat{\phi}$ and $\delta\hat{\dot{\phi}}$ are realizations of random fields with the same statistics as the quantum Minkowski vacuum fluctuations, and $\phi_{\rm fv}$ is the mean field value in the false vacuum.
Using these as initial conditions, we solve Hamilton's equations
\begin{equation}
  \frac{d\phi}{dt}       = \dot{\phi} \qquad \mathrm{and} \qquad
  \frac{d\dot{\phi}}{dt} = -\nabla^2\phi - V'(\phi) 
\end{equation}
with a tenth-order accurate Gauss-Legendre time integrator~\cite{Braden:2014cra,MR0159424}.
The spatial Laplacian is computed using a Fourier collocation stencil, resulting in periodic boundary conditions.
This combination conserves energy to $\mathcal{O}(10^{-15})$ with a maximal pointwise error of $\mathcal{O}(10^{-13})$ as either the grid spacing or time-step is varied.
Additionally, our time-integrator is symplectic so that phase space volume is preserved.
Finally, we verified that our numerical simulations are time-reversible.  First we integrate forward in time well into the regime where the false vacuum has decayed completely.  By reversing the flow of time and backward integrating for an equal time interval, we are able to recover the initial state of $\phi$ and $\dot{\phi}$ to a precision of $\mathcal{O}(10^{-15})$.
Further, if we continue to backward integrate in time, we again see the initial state decay since the false vacuum is not a true eigenstate of the Hamiltonian.

This approach does not capture quantum interference or the contributions of non-classical paths, which as we demonstrate in the supplementary material enter at $\mathcal{O}(\hbar^2)$ in our approach.
Nonetheless, there are many cases in which this semiclassical approach gives an accurate description of the statistics of the field(s).
Some well-known cases include post-inflation preheating dynamics and relativistic heavy ion collisions~\cite{Kofman:1994rk,Kofman:1997yn,Khlebnikov:1996mc,Felder:2000hq,Berges:2004ce}.
In these standard scenarios, individual Fourier modes become highly squeezed while in the linear regime, thus entering the classical wave limit and justifying a classical treatment of the field dynamics and statistics. 
The case considered here is novel, since once a bubble has formed in the medium the fluctuations are no longer statistically homogeneous, and the growth of modes leading to its formation may not occur in the linear regime.
However, a bubble is a highly classical field configuration, so we expect that many aspects will be properly captured by our simulations.
Regardless of how accurately these semiclassical simulations capture the full quantum statistics, they are more than sufficient to demonstrate the \emph{existence} of classically-allowed bubble nucleation trajectories.

\mySec{Classically Allowed Decays}
We now use the framework outlined above to study the dynamics of a scalar field $\phi$ initially trapped in a false vacuum.
Motivated by the analog false vacuum setup in cold atom Bose-Einstein condensate (BEC) systems~\cite{Fialko:2014xba,Fialko:2016ggg,Braden:2017add},
we consider the following action
\begin{equation}
  \label{eqn:lagrangian-scalar}
  \mathcal{L}_\phi = \frac{\dot{\phi}^2}{2} - \frac{(\nabla\phi)^2}{2} - V_0\left(-\cos\frac{\phi}{\phi_0} + \frac{\lambda^2}{2}\sin^2\frac{\phi}{\phi_0}\right) \, .
\end{equation}
In the BEC experiments, $\lambda$ is an experimentally tunable parameter controlling the depth of the false-vacuum well.
For $\lambda > 1$, this theory posesses an infinite sequence of false vacuum extrema at $\phi = (2k+1)\pi$ with $k\in \mathbb{Z}$.
Instead, for $\lambda < 1$ the false vacua are local maxima of the potential, and states centered around these points undergo spinodal decay.

Here, we approximate the initial vacuum fluctuations as those of a free massive scalar field with mass squared $m_{\rm eff}^2 = V''(\phi=\pi\phi_0) = V_0\phi_0^{-2}(-1+\lambda^2)$.
We sample fluctuations of the field $\delta\phi$ and its time derivative $\delta\dot{\phi}$ as uncorrelated Gaussian random fields with spectra
\begin{equation}
  \langle\phi_k^*\phi_{k'}\rangle = \frac{1}{2\omega_k}\delta(k-k') \qquad \langle\dot{\phi}_k^*\dot{\phi}_{k'}\rangle = \frac{\omega_k}{2}\delta(k-k') \, ,
\end{equation}
where $\omega_k^2 = k^2 + m_{\rm eff}^2$ and $\delta(k-k')$ is the Dirac delta function.

It is convenient to work with the dimensionless variable $\bar{\phi} = \phi/\phi_0$, along with dimensionless units $\bar{t} = \mu t$ and $\bar{x} = \mu x$ where $\mu$ is a constant with dimensions of mass.
Explicitly, for a one-dimensional discrete lattice of finite size $L$,
\begin{equation}
  \delta\hat{\bar{\phi}}({\bf x}) = \frac{1}{\phi_0} \frac{1}{\sqrt{2L}} \sum_{{\rm n}=1}^{{\rm n}_{\rm cut}} e^{i{\bf k}_{\rm n}\cdot {\bf x}} \frac{\hat{\alpha}_{\rm n}}{\left(m_{\rm eff}^2+k_{\rm n}^2\right)^{1/4}}  \, ,
\end{equation}
where $\hat{\alpha}_{\rm n}$ are realizations of complex Gaussian random deviates with unit variance, which we generate using $\hat{\alpha}_{\rm n} = \sqrt{-\ln\hat{a}_{\rm n}}e^{i2\pi\hat{b}_{\rm n}}$ where $\hat{a}_{\rm n}$ and $\hat{b}_{\rm n}$ are uniform random deviates on the unit interval.
We have included a sharp cutoff in the spectrum, which must be less than or equal to the Nyquist frequency and the wavevectors are given by $k_{\rm n} = \frac{2\pi {\rm n}}{L}$.
The $\delta\dot{\phi}$ fluctuations are initialized analogously.

\begin{figure}
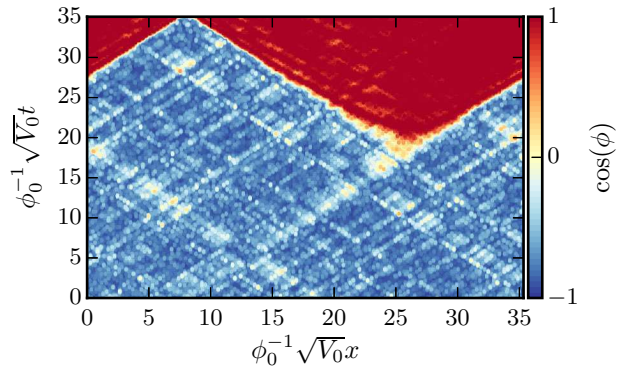

  \includegraphics[width=3.375in]{{{bubble-nucleation-trajectory}}}
  \caption{A classically-allowed false vacuum decay for the Lagrangian density~\eqref{eqn:lagrangian-scalar} with $\lambda = 1.2$ and $\phi_0 = \frac{\pi}{4\sqrt{2}}$.  The initial fluctuations $\delta\phi$ and $\delta\dot{\phi}$ are realizations of Gaussian random fields with spectra mimicking Minkowski false vacuum fluctuations.  We observe the dynamical emergence of a bubble with field localized near a true vacuum ($\cos\phi=1$) embedded within the false vacuum ($\cos\phi=-1$).  This explicitly demonstrates the existence of classically-allowed paths for false vacuum decay.}
  \label{fig:nucleation-scalar}
\end{figure}

\Figref{fig:nucleation-scalar} shows an example classical field evolution in one spatial dimension for a single initial field realization described above.
We see a bubble (\ie domain wall-antiwall pair) nucleate from the initial vacuum fluctuations, then subsequently expand into the surrounding false vacuum.
To our knowledge, this is the first explicit demonstration that bubble nucleation in false vacuum decay \emph{can} occur via classically-allowed field evolutions in pure relativistic field theory.
Aside from the nature of quantum vs.\ classical statistics, this is analogous to bubble formation in thermal~\cite{Linde:1981zj} (or other classical statistical~\cite{Khlebnikov:1998sz}) phase transitions.
Thermal nucleation events are also typically modeled as random, but at a microscopic level the full set of field fluctuations (including the ``thermal bath'') are simply following the classical equations of motion.
Bubble nucleations are special classical trajectories through phase space that describe the coherent creation of a bubble.
In the quantum case we are modeling, the fluctuations instead descend directly from the initial state of the field rather than an external bath.

Let us now consider how to interpret the results of our simulations.
A complete time-dependent description requires obtaining the wave functional
\begin{equation}
  \langle\phi({\bf x})|\Psi(t)\rangle = \int\mathcal{D}\phi_i \langle\phi|\hat{U}(t|t_0)|\phi_i\rangle\langle\phi_i({\bf x})|\Psi(t_0)\rangle
\end{equation}
for the full scalar field configuration as a function of time, given the initial wave-functional $|\Psi(t_0)\rangle$.
However, any observer watching the false vacuum boil will not have access to the full quantum state of the field as encoded in $|\Psi(t)\rangle$.
Rather than seeing a quantum superposition of nucleation events, 
such an observer will see $N_b$ bubbles nucleate with spacetime locations $\{({\bf x}_i,t_i),i=1,N_b\}$.
A priori, we can only predict probabilities for $N_b$ and these spacetime positions.
Given that an actual set of nucleation events has taken place, the observer should be able to describe the field evolution subject to the constraint of these observed nucleations.
Since the observer will not have perfect knowledge of the multi-bubble state, their observations should be marginalized over fluctuations around the constrained field path.
The result presented here shows that there exist constrained paths describing purely classical time-evolution connecting final multi-bubble states directly to realizations of the initial vacuum fluctuations.
Since these classical paths extremize the action, we expect that they will provide an important contribution to the propagator determining the evolution of the wave-functional.

To contrast our approach, let us instead briefly consider the standard Euclidean (\ie instanton) formalism~\cite{Coleman:1977py,Callan:1977pt,Weinberg:2012pjx}.
We Wick-rotate to Euclidean time, then search for extremal paths of the Euclidean action connecting the false vacuum at infinity to the true vacuum at the origin.
These extremal paths are known as bounce solutions.
Nearly all existing treatments assume the bounce is spherically-symmetric in Euclidean signature.
This dramatically reduces the phase space of allowed field configurations, simplifying it to a quantum mechanics problem.
Nearly all existing treatments assume the bounce is spherically-symmetric.
In Lorentzian signature, the bounce describes a spherically-symmetric bubble that contracts from infinite size in the distant past, reaches a minimum turnaround radius, and then re-expands.
The nucleation event itself is modeled by artificially removing the contracting phase, then pasting the remaining expanding phase into the ambient spacetime.
The pasting step is interpreted as a classically-forbidden quantum tunneling process, analogous to the decay of an unstable particle in quantum mechanics.

However, the tunneling interpretation is fueled by analogy with quantum mechanics, which as mentioned above requires a drastic reduction in the allowed phase space for the field evolution.
We have demonstrated that the neglected fluctuations can play a pivotal role around the time of nucleation, and allow for bubbles to nucleate via classically-allowed trajectories.
Specific fluctuation configurations that may lead to bubble formation are: (1)~local peaks or valleys in the initial fluctuations around the false vacuum that probe the nonlinear structure of the potential and act as bubble nucleation seeds, or (2)~the repeated mildly nonlinear interactions of propagating waves that eventually lead to the development of a local field peak from which a bubble may form.  The latter effect means that energy within a small region may not be conserved due to energy transport via currents.

\mySec{False Vacuum Decay Rates}
Thus far, we showed the existence of classically-allowed paths through field space describing false vacuum decay via bubble nucleation.
Given such paths exist, we expect they play an important part in the false vacuum decay process.
They may either be a competing process with classically-forbidden bubble nucleations;
or, given the drastic dimensional reduction used in the Euclidean formalism, 
may actually be a more complete description of the decays captured by instanton techniques.
To investigate this issue, we estimate the bubble nucleation rate using our real-time simulations and compare it with the instanton predictions.

In $D=d+1$ spacetime dimensions, the Euclidean prediction for the nucleation rate per unit volume is
\begin{eqnarray}\label{eqn:decay-rate-inst}
  \frac{\Gamma}{V} &=& \left(\frac{B}{2\pi}\right)^{D/2}e^{-B}\left(\frac{\mathrm{det'  \delta^2 S_{\rm E}[\phi_{\rm B}]}}{\mathrm{det}\delta^2S_{\rm E}[\phi_{\rm fv}]}\right)^{-1/2} \nonumber \\
  &\equiv& \left(\frac{B}{2\pi}\right)^{D/2}e^{-B} D(\lambda,\phi_0) \, ,
\end{eqnarray}
where $B = S_{\rm E}[\phi_{\rm B}] - S_{\rm E}[\phi_{\rm fv}]$ is the difference in the Euclidean action evaluated on the bounce solution and in the false vacuum, $\mathrm{det}'$ indicates the determinant with zero modes removed, and $\delta^2S_{\rm E}$ is the second variation of the Euclidean action.
The bounce $\phi_{\rm B}$ satisfies
\begin{equation}
    \frac{\partial^2\phi_{\rm B}}{\partial r_{\rm E}^2} + \frac{d}{r_{\rm E}}\frac{\partial \phi_{\rm B}}{\partial r_{\rm E}} - \frac{\partial V}{\partial \phi} = 0 \, ,
\end{equation}
with boundary conditions $\phi_{\rm B}(r_{\rm E}=\infty) = \phi_{\rm fv}$ and $\partial_{r_{\rm E}}\phi(r_{\rm E}=0) = 0$.  In addition, $\phi_{\rm B}(r_{\rm E}=0)$ should be near the true vacuum, and the solution should possess a single negative eigenmode to  give an imaginary contribution to the Euclidean effective action~\cite{Coleman:1987rm}.
A precise calculation of the determinant prefactor is beyond the scope of this paper, so we parametrize it as $D(\lambda,\phi_0) = g(\lambda,\phi_0)\left(V_0\phi_0^{-2}\right)^{D/2}$, where $g$ is a dimensionless function of order unity which includes counterterms to renormalize the determinant.\footnote{When the bounce solution possesses a hierarchy of scales, such as in the thin-wall limit, $g$ may not be order one.}
For the purpose of obtaining scaling relations, it is convenient to re-express~\eqref{eqn:decay-rate-inst} as
\begin{equation}\label{eqn:decay-rate-nodim}
  \frac{\bar{\Gamma}}{\bar{L}} \equiv \frac{1}{\mu^2}\frac{\Gamma}{L} = 2g(\lambda,\phi_0)\left(\frac{V_0}{\mu^2\phi_0^2}\right)\phi_0^2C(\lambda)e^{-2\pi\phi_0^2C(\lambda)} \, ,
\end{equation}
where
$C(\lambda) = \int \bar{r}_{\rm E}d\bar{r}_{\rm E} \left(\frac{(\partial_{\bar{r}_{\rm E}}\bar{\phi}_{\rm B})^2}{2} + \frac{1}{\mu^2\phi_0^2}\left[V(\phi)-V(\phi_{\rm fv})\right]\right)$
and we have specialized to one spatial dimension.
For our particular model, 
$V(\phi) = V_0\left(-\cos\phi + \frac{\lambda^2}{2}\sin^2\phi\right)$.
The factor of $2$ accounts for the independent instanton decay paths in this model.

As is clear from~\figref{fig:nucleation-scalar}, the emergence of a bubble is a ``fuzzy'' process which makes a precise determination of the nucleation time somewhat ambiguous.
We therefore adopt the following operational definition.
Imagine an ensemble of observers each watching a finite region of the false vacuum boil.
We ask each of these observers to provide us with a history of $\langle\cos\phi\rangle_{\rm V}$ within their region, where $\langle\cdot\rangle_{\rm V}$ represents a spatial average.
To obtain statistics, we make one additional assumption --- each field trajectory (and thus each observer) is weighted by the probability of the initial realization of the fluctuations.
In the particular case considered here, where we assume the initial state is Gaussian and draw samples from an ensemble of Gaussian random fields, each simulation is weighted equally.
We select a threshold, and declare that the observer's region has decayed at the first passage of $\langle\cos\phi\rangle_{\rm V}$ past this threshold.
To reduce the effects of multiple bubbles and fluctuations, we choose the threshold to be $\bar{c}_{\rm T}+ n_{\sigma}\Delta c_{\rm T}$ for some constant $n_{\sigma}$.
  Here $\bar{c}_{\rm T}$ and $\Delta c_{\rm T}$ are, respectively, the ensemble average and standard deviation of $\langle\cos\phi\rangle_{\rm V}$ on the initial time slice.
  We then run ensembles of $5000$ lattice simulations with varying $\phi_0$ and side length $L$ for $\lambda = 1.2$, and empirically determine the fraction of undecayed trajectories $F_{\rm survive}(t)$.
We verified $F_{\rm survive}$ is insensitive to the choices of both the lattice spacing $dx$ and the threshold $n_{\sigma}$ for $5 < n_{\sigma} < 25$.
After an initial transient that depends on the simulation volume $L$, $F_{\rm survive}$ enters a regime of nearly exponential decay.  We fit this exponential tail $F_{\rm survive}= e^{-\Gamma(t-t_o)}$ with parameters $\Gamma$ and $t_0$ to extract the decay rate and verified that $\frac{\Gamma}{L}$ is independent of the simulation volume $L$.

\begin{figure}
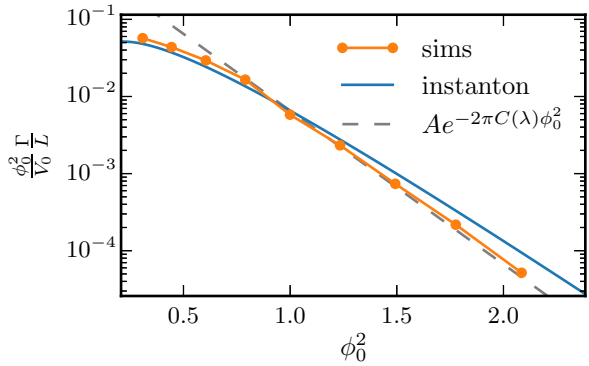

  \includegraphics[width=3.375in]{{{decay-rates}}}
  \caption{Decay rates of false vacuum initial states as $\phi_0$ is varied for our ensembles of lattice simulations (orange dots).  For comparison, the blue solid line is the analytic prediction~\eqref{eqn:decay-rate-nodim} with $g = -1+\lambda^2 = 0.44$ to set the overall scale to the false vacuum mass, and the gray dashed line is the leading-order prediction (\ie $\Gamma = Ae^{-2\pi C(\lambda)\phi_0^2}$ with the constant $A$ normalized so the curves match at $\phi_0=1$).  We fixed $\lambda = 1.2$ and $\frac{V_0}{\phi_0^2\mu^2} = 8\times 10^{-3}$.}
  \label{fig:decay-rates}
\end{figure}
In~\figref{fig:decay-rates} we compare the decay rates extracted from our simulations with those of the instanton formalism (obtained using the method of Ref.~\cite{Bond:2015zfa}).
We also include the leading order prediction to compare to an exponential falloff at large $\phi_0$.
To capture all relevant dynamical modes, we set the spectral cutoff $k_{\rm cut}=\frac{1024\pi}{25\sqrt{2}}\frac{\sqrt{V_0}}{\phi_0}$ above the wavenumber where the Fourier amplitude of the bounce profile $\phi_{\rm B}$ dropped below $10^{-15}$.  In this regime, increasing $k_{\rm cut}$ should simply renormalize the field theory.
Given the unknown determinant prefactor, exponential sensitivity of the decay rate to the bounce action, and multi-bubble and fluctuation effects entering into our decay time extraction, our simulations match the instanton prediction very well.
  In particular, the amplitude is the correct order of magnitude, and the rate of exponential decay at large $\phi_0$ closely matches the Euclidean calculation.
  
We also varied $\lambda$ at fixed $\phi_0^2C(\lambda)$ and found the time-evolution of the survival fraction to be nearly independent of the choice of $\lambda$.
This suggests not only that classically-allowed decay paths are a key process in false vacuum decay, but also that these classically-allowed paths may be the same ones being captured by the standard instanton calculation.
In order to address this latter issue, the results presented here motivate a deeper investigation of the function $g$, renormalization effects, and a more sophisticated bubble extraction algorithm.

\mySec{Conclusions}
We outlined a novel real-time framework to study quantum false vacuum decay, based on semi-classical stochastic lattice simulations.
This provides a step towards incorporating the full temporal evolution of a decaying region of the false vacuum.
In particular, we demonstrated that the false vacuum can decay via bubble nucleation along classically-allowed field trajectories, and the decay rates of such decays agree well with standard computations based on the instanton formalism.
This suggests that our formalism provides a description of the ``fuzzy'' process of a bubble nucleating in false vacuum.
We will explore how our interpretation generalizes to other systems of non-equilibrium bubble formation in a forthcoming work.

One particularly exciting prospect is the possiblity of emulating the false vacuum decay process with cold atom systems, suggesting we may be able to experimentally test the picture outlined here~\cite{Fialko:2014xba,Fialko:2016ggg,Braden:2017add}.
In addition, our approach opens new avenues of investigation which cannot be addressed within the Euclidean framework.  These include decay rates from non-vacuum initial states, investigation of bubble-bubble correlations, and evolution in a regime of rapid decays.
Furthermore, the emergence of a classical bubble from a highly quantum initial state is an example of self-decoherence of a quantum field, and may yield insight into the process by which effectively classical density perturbations arise from quantum fluctuations during inflation~\cite{Polarski:1995jg,Kiefer:1998qe,Burgess:2014eoa,Nelson:2016kjm}.
Finally, the emergence of a classical bubble from the quantum vacuum bears many similarities to the process of ``measuring'' a bubble nucleation event, and may shed light on the notion of measurement in quantum field theory and quantum cosmology~\cite{Zurek:2003zz}.

\section*{Acknowledgments}
\begin{acknowledgments}
  JB and HVP were supported by the European Research Council (ERC) under the European Community's Seventh Framework Programme (FP7/2007-2013)/ERC grant agreement number 306478- CosmicDawn.  MCJ was supported by the National Science and Engineering Research Council through a Discovery grant.  AP was supported by the Royal Society.  SW acknowledges financial support provided under the Royal Society University Research Fellow (UF120112), the Nottingham Advanced Research Fellow (A2RHS2), the Royal Society Project (RG130377) grants, and the EPSRC Project Grant (EP/P00637X/1). This work was partially enabled by funding from the UCL Cosmoparticle Initiative.  This research was supported in part by Perimeter Institute for Theoretical Physics. Research at Perimeter Institute is supported by the Government of Canada through the Department of Innovation, Science and Economic Development Canada and by the Province of Ontario through the Ministry of Research, Innovation and Science. 
\end{acknowledgments}

\bibliography{classical-decay}{}

\onecolumngrid

\section{Supplemental Material: Semiclassical Statistical Simulations as Evolution of the Wigner Functional}

In this supplementary material, we demonstrate that the statistical lattice simulations employed in the main text is equivalent to quantum dynamics to first order in an expansion in $\hbar$. A similar derivation can be found in Ref.~\cite{Mrowczynski:1994nf}

Consider the Wigner functional
\begin{equation}
  W[\phi,\Pi] = \int\mathcal{D}\eta\, e^{-\frac{i}{\hbar}\int{\rm d}^dy\, \Pi(y)\eta(y)} \left\langle\phi+\frac{\eta}{2}\bigg| \hat{\rho}\bigg| \phi-\frac{\eta}{2}\right\rangle
\end{equation}
which is normalized as
\begin{equation}
  \int \mathcal{D}\phi\mathcal{D}\Pi \,\, W[\phi,\Pi] = 1 \, ,
\end{equation}
where we have absorbed some factors of $2\pi$ into the definition of $\mathcal{D}\Pi$.
Using $W$, expectation values of an operator $\hat{\mathcal{O}}(\hat{\phi},\hat{\Pi})$ are obtained from
\begin{equation}
  \langle\hat{\mathcal{O}}(\hat{\phi},\hat{\Pi})\rangle = \int\mathcal{D}\phi\mathcal{D}\Pi \,\, W[\phi,\Pi]\mathcal{O}_{\rm W}(\phi,\Pi) 
\end{equation}
where $\mathcal{O}_{\rm W}(\phi,\Pi)$ is the Wigner symbol of $\mathcal{O}$.
For a fully symmetrized operator, $\mathcal{O}_{\rm W}$ is obtained by simply replacing the quantum operators $\hat{\phi}$ and $\hat{\Pi}$ by classical functions.
Thus, the Wigner functional acts as a quantum probability distribution on phase space, although one should bear in mind that it is not always positive definite.  However, for a Gaussian state, the Wigner function is positive definite and the probability interpretation holds.

Consider a theory with Hamiltonian density of the form
\begin{equation}
  H = \int {\rm d}^dx\, \mathcal{H} = \int {\rm d}^dx\, \left[\frac{\Pi^2}{2} + \frac{(\nabla\phi)^2}{2} + V(\phi)\right] \, .
\end{equation}
Working in the Schrodinger picture, the equation of motion for the density matrix $\hat{\rho}$ is
\begin{equation}
  \frac{d\hat{\rho}}{dt} = \frac{1}{i\hbar}\left[\hat{H},\hat{\rho}\right]\, .
\end{equation}
Using this, we can derive the evolution equation for the Wigner functional
\begin{equation}
    \left[\frac{\partial}{\partial t} + \int d^dx\left(\Pi\frac{\delta}{\delta\phi} + \nabla^2\phi\frac{\delta}{\delta \Pi} - \frac{2}{i\hbar}V(\phi)\sin\left(\frac{i\hbar}{2}\overleftarrow{\partial}_\phi\frac{\delta}{\delta\Pi}\right)\right)\right]W[\phi,\Pi;t] = 0 \,
\end{equation}
where $\overleftarrow{\partial}_\phi$ acts only on the potential and $\delta/\delta\Pi$ acts on the Wigner functional.
Expanding the sine term arising from the potential in $\hbar$, we obtain
\begin{equation}
    \left[\frac{\partial}{\partial t} + \int d^dx\left(\frac{\delta\mathcal{H}}{\delta\Pi}\frac{\delta}{\delta\phi} - \frac{\delta\mathcal{H}}{\delta\phi}\frac{\delta}{\delta \Pi} + \mathcal{O}\left(\hbar^2V'''(\phi)\frac{\delta^3}{\delta\Pi^3}\right)\right)\right] W[\phi,\Pi;t] = 0 \, .
\end{equation}
In the above, we have used the classical equations of motion
\begin{subequations}
\begin{align}
  \frac{\mathrm{d}\phi}{\mathrm{d}t} &= \frac{\delta\mathcal{H}}{\delta\Pi} = \Pi \\
  \frac{\mathrm{d}\Pi}{\mathrm{d}t} &= -\frac{\delta\mathcal{H}}{\delta\phi} = \nabla^2\phi - V'(\phi) \, .
\end{align}
\end{subequations}
Therefore, to $\mathcal{O}(\hbar^2)$ in the dynamical evolution, we can solve for $W$ using the method of characteristics, with each characteristic satisfying the classical equations of motion from some initial $\phi$ and $\Pi$ configuration.
This cleanly factorizes the effects of quantum mechanics into: (1) a contribution from the initial fluctuation statistics encoded in $W[\phi,\Pi,t=0]$, and (2) $\mathcal{O}(\hbar^2)$ and higher modifications to the classical dynamical evolution.
Our approach thus corresponds to this approximation, with the initial state approximated as a Gaussian.
Since the initial fluctuation spectra are set by commutation relations between $\phi$ and $\Pi$, in this formalism the leading quantum correction in $\hbar$ arises from the initial form of $W$, rather than due to dynamical evolution.
Note the crucial role the existence of classically-allowed false vacuum decay paths plays in the utility of this approach.

When considering Fig.~2, note that the instanton calculation is also an expansion in $\hbar$. The leading effect arises from the action on the instanton trajectory $e^{-S_{\rm I}}$, while the fluctuation determinant gives the first quantum correction (including renormalization effects).
Thus, both the curves are approximations to the true tunneling rate and a careful calculation of the determinant prefactor could bring them into even better agreement. 

\end{document}